\documentclass[a4paper,10pt,runningheads]{llncs}
\usepackage{url}
\usepackage{bbm}
\usepackage{amsmath}
\usepackage{amssymb}
\usepackage{upgreek}
\usepackage[mathletters]{ucs}
\usepackage[utf8x]{inputenx}
\usepackage{comment}
\newcommand{\N}{\ensuremath{\mathbbm{N}}}
\newcommand{\Z}{\ensuremath{\mathbbm{Z}}}
\newcommand{\Q}{\ensuremath{\mathbbm{Q}}}
\usepackage{wrapfig}
\usepackage{color}

\usepackage{graphicx}

\definecolor{dkblue}{rgb}{0,0.1,0.5}
\definecolor{lightblue}{rgb}{0,0.5,0.5}
\definecolor{dkgreen}{rgb}{0,0.4,0}
\definecolor{dk2green}{rgb}{0.4,0,0}
\definecolor{dkviolet}{rgb}{0.6,0,0.8}

\usepackage{listings}

\begin{document}
\title{Type Classes for Mathematics in Type Theory}
\author{Bas Spitters \and Eelis van der Weegen} 
\institute{Radboud University Nijmegen}
\date{today}
\maketitle
\begin{abstract}
The introduction of first-class type classes in the Coq system calls for re-examination of the basic interfaces used for mathematical formalization in type theory.
We present a new set of type classes for mathematics and take full advantage of their unique features to make practical a particularly flexible approach formerly thought infeasible. Thus, we address both traditional proof engineering challenges as well as new ones resulting from our ambition to build upon this development a library of constructive analysis in which abstraction penalties inhibiting efficient computation are reduced to a minimum.

The base of our development consists of type classes representing a standard algebraic hierarchy, as well as portions of category theory and universal algebra. On this foundation we build a set of mathematically sound abstract interfaces for different kinds of numbers, succinctly expressed using categorical language and universal algebra constructions. Strategic use of type classes lets us support these high-level theory-friendly definitions while still enabling efficient implementations unhindered by gratuitous indirection, conversion or projection.

Algebra thrives on the interplay between syntax and semantics. The Prolog-like abilities of type class instance resolution allow us to conveniently define a quote function, thus facilitating the use of reflective techniques.
\end{abstract}

\section{Introduction}
The development of libraries for formalized mathematics presents many software engineering challenges~\cite{C-corn,DBLP:conf/types/HaftmannW08}, because it is far from obvious how the clean, idealized concepts from everyday mathematics should be represented using the facilities provided by concrete theorem provers and their formalisms, in a way that is both mathematically faithful and convenient to work with.

For the algebraic hierarchy---a critical component in any library of formalized mathematics---these challenges include structure inference, handling of multiple inheritance, equality of axiomatically posited and derived structure, idiomatic use of notations, support for models based on quotient representations, and convenient algebraic manipulation (e.g.\ rewriting). Several solutions have been proposed for the Coq theorem prover: dependent records~\cite{DBLP:journals/jsc/GeuversPWZ02} (a.k.a. telescopes), packed classes~\cite{Packed}, and occasionally modules. We present a new solution based entirely on the use of Coq's new type class facility to make fully `unbundled' predicate-representations of algebraic structures practical to work with.

Our development is not merely aimed at formalization of theory, and our choice of a system based on type theory is no accident. It is our explicit ambition for the interfaces and theory we develop to be employed directly for the specification and parameterization of efficiently executable procedures and data structures, implemented using type theory's native term reduction as a programming language. Thus, our work belongs in the long tradition of realize the promise of type theory to truly unite mathematical formalization and certified (functional) programming, without making painful sacrifices on either side.

Because our ``ultimate'' goal is to use this development as a basis for constructive analysis with practical certified exact real arithmetic, and because numerical structures are ideal test subjects for our algebraic hierarchy, we shall use these to motivate and demonstrate the key parts of our development. Since we are concerned with \emph{efficient} computation, we want to be able to effortlessly swap implementations of number representations. Doing this requires that we have clean abstract interfaces, and mathematics tells us what these should look like: we represent $\N$, $\Z$, and $\Q$ as \emph{interfaces} specifying an initial semiring, an initial ring, and a field of integral fractions, respectively. To express these interfaces elegantly and without duplication, our development\footnote{The sources are available at:~\url{http://www.eelis.net/research/math-classes/}} includes an integrated formalization of parts of category theory and multi-sorted universal algebra, all expressed using type classes for optimum effect.

In this paper we focus on the Coq proof assistant. We conjecture that the methods can be transferred
to any type theory based proof assistant supporting type classes such as
Matita~\cite{asperti2007user}.

\emph{Outline.}
In section~\ref{preliminaries} we briefly describe the Coq system and its implementation of type classes. Then, in section~\ref{bundling}, we give a very concrete introduction to the issue of \emph{bundling}, arguably the biggest design dimension when building interfaces for abstract structures. In section~\ref{predicateclasses} we show how type classes can make the use of `unbundled' purely predicate based interfaces for abstract structures practical.

In the rest of the paper, we tour the key components in our development, leading up to the numerical interfaces. This will not only show the pleasant style of formalization that rigorous use of type classes enables, but will also show that eager adoption and incorporation of more abstract mathematical perspectives (which are traditionally often ignored when doing dependently typed programming on concrete data structures in type theory), is not only feasible but actually practical and beneficial.

 In section~\ref{classes}, we discuss our algebraic hierarchy implemented with type classes. In sections~\ref{cat} and~\ref{univ} we give a taste of what category theory and universal algebra look like in our development, and in section~\ref{numbers} we use these facilities to build abstract interfaces for numbers. In order to illustrate a very different use of type classes, we discuss the implementation of a quoting function for algebraic terms in terms of type classes, in section~\ref{quoting}. In section~\ref{sequences} we hint at an interface for sequences, but describe how a limitation in the current implementation of Coq makes its use problematic. We end with conclusions in section~\ref{conclusions}.

\section{Preliminaries}
\label{preliminaries}

The Coq proof assistant is based on the calculus of inductive
constructions~\cite{CoquandHuet,CoquandPaulin}, a dependent type theory with (co)inductive types; see~\cite{BC04,Coq}. In true Curry-Howard fashion, it is both an excessively pure, if somewhat pedantic, functional programming language with an extremely expressive type system, and a language for mathematical statements and proofs. We highlight some aspect of Coq that are of particular relevance to our development.

\paragraph{Types and propositions.}

Propositions in Coq are types~\cite{CMCP,ITT}, which themselves have types called \emph{sorts}. Coq features a distinguished sort called \lstinline|Prop| that one may choose to use as the sort for types representing propositions. The distinguishing feature of the \lstinline|Prop| sort is that terms of non-\lstinline|Prop| type may not depend on the values of inhabitants of \lstinline|Prop| types (that is, proof terms).
This regime of discrimination establishes a weak form of proof irrelevance, in that changing a proof can never affect the result of value computations. On a very practical level, this lets Coq safely erase all \lstinline|Prop| components when extracting certified programs to OCaml or Haskell.

Occasionally there is some ambiguity as to whether a certain piece of information (such as a witness to an existential statement) is strictly `proof matter' (and thus belongs in the \lstinline|Prop| sort) or actually of further computational interest (and thus does \emph{not} belong to the \lstinline|Prop| sort). We will see one such case when we discuss the first homomorphism theorem in section \ref{homothm}. Coq provides a modest level of \emph{universe-polymorphism} so that we may avoid duplication when trying to support \lstinline|Prop|-sorted and non-\lstinline|Prop|-sorted content with a single set of definitions.

\paragraph{Equality, setoids, and rewriting}

The `native' notion of equality in Coq is that of term convertibility, naturally reified as a proposition by the inductive type family \lstinline|eq: ∀ (T: Type), T → T → Prop| with single constructor \lstinline|eq_refl|:
\begin{lstlisting}
  eq_refl : ∀ (T: Type) (x: T), x ≡ x,
\end{lstlisting}
where `\lstinline|a ≡ b|' is notation for \lstinline|eq T a b|. Here we diverge from Coq tradition and reserve the `\lstinline|a = b|' notation for \emph{setoid} equality (to be discussed momentarily), as this is the equality we will be working with most of the time.

Importantly, since convertibility is a congruence, a proof of \lstinline|a ≡ b| lets us substitute \lstinline|b| for \lstinline|a| anywhere inside a term without further conditions. We mention this explicitly only because such rewriting \emph{does} give rise to conditions when we depart from raw convertibility and introduce equivalence relations that express how possibly distinct (unconvertible) terms may represent the same conceptual object. Rational numbers represented by (non-reduced) formal integer fractions are a typical example. Rewriting a subterm using a proof of such an equality is permitted only if the subterm is argument to a function that has been proven to \emph{respect} the equality. Such a function is called \emph{proper} with respect to the equality in question, and propriety must be proved for each function in whose arguments we wish to enable rewriting.

Because the Coq type theory lacks quotient types (as it would make type checking undecidable), one usually bases abstract structures on a \emph{setoid} (`Bishop set'): a type equipped with an equivalence relation~\cite{Bishop67,Hofmann,Capretta}. Palmgren~\cite{palmgren2009constructivist} shows that Bishop sets have pleasant categorical properties, which translate to a powerful implicit type structure. It would be of interest to actually provide machine support for this type structure. As we will see in section~\ref{univ}, working with setoids pays off when working with notions such as quotient algebras.

Effectively keeping track of, resolving, and combining proofs of equivalence-ness and propriety when the user attempts to substitute a given (sub)term using a given equality, is known as ``setoid rewriting'', and requires nontrivial infrastructure and support from the system. The Coq implementation of these mechanisms was largely rewritten by Matthieu Sozeau~\cite{Setoid-rewrite} to make it more flexible and to replace the old special-purpose setoid/morphism registration command with a clean type class based interface.

The algebraic hierarchy of the \textsc{Ssreflect} libraries~\cite{Packed} uses an alternative approach. It simply requires canonical representation of all objects, so that setoid equality is not needed. Of course, this policy severely restricts the freedom one has when implementing models of abstract structures. Indeed, for some sets, canonical representation schemes do not exist. The constructive reals, which are of particular interest to us, are an example of such a set.

\paragraph{Type classes.}

Type classes \cite{wadler1989adhoc} have been a great success story in the Haskell functional programming language, as a means of organizing interfaces of abstract structures. Coq's type classes provide a superset of their functionality, but implemented in a different way.

In Haskell and Isabelle, type classes and their instances are second class. They are handled as specialized syntactic constructs whose semantics are given specifically by the type class apparatus. By contrast, the expressivity of dependent types and inductive families as supported in Coq, combined with the use of pre-existing technology in the system (namely proof search and implicit arguments) enable a \emph{first class} type class implementation~\cite{DBLP:conf/tphol/SozeauO08}: classes are ordinary record types (`dictionaries'), instances are ordinary constants of these record types (registered as \emph{hints} with the proof search machinery), class constraints are ordinary implicit parameters, and instance resolution is achieved by augmenting the unification algorithm to invoke ordinary proof search for implicit arguments of class type.
Thus, type classes in Coq are realized by relatively minor syntactic aids that bring together existing facilities of the theory and the system into a coherent idiom, rather than by introduction of a new category of qualitatively different definitions with their own dedicated semantics.

The basic idea of using type-class-like facilities for structuring computerized mathematics dates back to the \textsc{AXIOM} computer algebra system~\cite{jenks1992axiom}. Weber~\cite{weber1993type} pursues the analogy between \textsc{AXIOM}'s so-called categories and type classes in Haskell. Santas~\cite{santas1995type} pursues analogies between type classes, \textsc{AXIOM} categories and existential types. Existential types are present in Haskell, but absent from Coq.

\section{Bundling is bad}\label{bundling}

Algebraic structures are expressed in terms of a number of carrier sets, a number of operations and relations on these carriers, and a number of laws that the operations and relations satisfy. In a system like Coq, we have different options when it comes to representing the grouping of these components. On one end of the spectrum, we can simply define the (conjunction of) laws as an $n$-ary predicate over $n$ components, forgoing explicit grouping altogether. For instance, for the mundane example of a reflexive relation, we could use:

\begin{lstlisting}
  Definition reflexive {A: Type} (R: relation A): Prop := ∀ a, R a a.
\end{lstlisting}
The curly brackets used for \lstinline|A| mark it as an implicit argument.

More elaborate structures, too, can be expressed as predicates (expressing laws) over a number of carriers, relations, and operations. While optimally flexible in principle, in practice \emph{naive} adoption of this approach (that is, without using type classes) leads to substantial inconveniences in actual use: when \emph{stating} theorems about abstract instances of such structures, one must enumerate all components along with the structure (predicate) of interest. And when \emph{applying} such theorems, one must either enumerate any non-inferrable components, or let the system spawn awkward metavariables to be resolved at a later time. Importantly, this also hinders proof search for proofs of the structure predicates, making any nontrivial use of theorems a laborious experience. Finally, the lack of \emph{canonical names} for particular components of abstract structures makes it impossible to associate idiomatic notations with them.

In the absence of type classes, these are all very real problems, and for this reason the two largest formalizations of abstract algebraic structures in Coq today, CoRN~\cite{C-corn} and \textsc{Ssreflect}~\cite{Packed}, both use \emph{bundled} representation schemes, using records with one or more of the components as fields instead of parameters. For reflexive relations, the following is a fully bundled representation---the other end of the spectrum:

\begin{lstlisting}
  Record ReflexiveRelation: Type :=
    { rr_carrier: Type
    ; rr_rel: relation rr_carrier
    ; rr_proof: ∀ x, rr_rel x x }.
\end{lstlisting}
Superficially, this instantly solves the problems described above: reflexive relations can now be declared and passed as self-contained packages, and the \lstinline|rr_rel| projection now constitutes a canonical name for relations that are known to be reflexive, and we could bind a notation to it. While there is no conventional notation for reflexive relations, the situation is the same in the context of, say, a semiring, where we would bind $+$ and $*$ notations to the record field projections for the addition and multiplication operations, respectively.

Unfortunately, despite its apparent virtues, the bundled representation introduces serious problems of its own, the most immediate and prominent one being a lack of support for \emph{sharing} components between structures, which is needed to cope with overlapping multiple inheritance.

In our example, lack of sharing support rears its head as soon as we try to define \lstinline|EquivalenceRelation| in terms of \lstinline|ReflexiveRelation| and its hypothetical siblings bundling symmetric and transitive relations. There, we would need some way to make sure that when we `inherit' \lstinline|ReflexiveRelation|, \lstinline|SymmetricRelation|, and \lstinline|TransitiveRelation| by adding them as fields in our bundled record, they all refer to the same carrier and relation. Adding additional fields stating equalities between the three bundled carriers and relations is neither easily accomplished (because one would need to work with heterogenous equality) nor would it permit natural use of the resulting structure (because one would constantly have to rewrite things back and forth).

Manifest fields~\cite{Pollack:2002} have been proposed to address exactly this problem. In fact, a semblance has been implemented in the Matita system~\cite{sacerdoti2008working}. We hope to convince the reader that type system extensions of this nature, designed to mitigate particular symptoms of the bundled approach, are less elegant than a solution (described in the next section) that avoids the problem altogether by using predicate-like type classes in place of bundled records.

If we revert back to the predicate formulation of relations, we \emph{could} still define \lstinline|EquivalenceRelation| in a bundled fashion without the need for equalities:
\begin{lstlisting}
  Record EquivalenceRelation: Type :=
    { er_carrier: Type
    ; er_rel: relation er_carrier
    ; er_refl: ReflexiveRelation er_carrier er_rel
    ; er_sym: SymmetricRelation er_carrier er_rel
    ; er_trans: TransitiveRelation er_trans er_rel }.
\end{lstlisting}
However, as before we conclude that \mbox{\lstinline|EquivalenceRelation|,} too, should be a predicate. Indeed, it would be rather strange for the interface of equivalence relations to differ qualitatively from the interface of reflexive relations.

Another attempt to recover some grouping might be to bundle the carrier with the relation into a (lawless) record, but this too hinders sharing. As soon as we try to define an algebraic structure with two reflexive relations on the same carrier, we need awkward hacks to establish equality between the carrier projections of two different (carrier, relation) bundles.

Even bundling just the operations of an algebraic structure together in a record (with the carrier as a parameter) leads to the same problem when, for example, one attempts to define a hypothetical algebraic structure with two binary relations and a constant such that both binary relations form a monoid with the constant.

A second problem with bundling is that as the bundled records are stacked to represent higher and higher structures, the projection paths for their components grow longer and longer, resulting in ever more unwieldy terms (coercions and notations can make this less painful). Further, if one tries to implement some semblance of sharing in a bundled representation, these projection paths additionally become non-canonical, and still more extensions have been proposed to address this symptom, e.g.\ coercion pullbacks~\cite{Hints}.

Thus, bundled representations come at a substantial cost in flexibility. Historically, using bundled representations has nevertheless been an acceptable trade-off, because (1) the unbundled alternative was such a pain, and (2) the standard algebraic hierarchy (up to, say, fields and modules) is not all that wild.

In the next section, we show that type-classification of structure predicates and their component parameters has the potential to remedy the problems associated with the naive unbundled predicate approach.

The observant reader may wonder whether it might be beneficial to go one step further and unbundle proofs of laws and inherited substructures as well. This is not the case, because there is no point in sharing them. After all, by (weak) proof irrelevance, the `value' of such proofs can be of no consequence anyway. Indeed, parameterizing on proofs would be actively harmful because instantiations differing only in the proof argument would express the same thing yet be nonconvertible, requiring awkward conversions and hindering automation.

\section{Predicate classes and operational classes}\label{predicateclasses}

To show that the fully unbundled approach with structures represented by predicates can be made feasible using type classes, we will tackle one by one the problems traditionally associated with their use, starting with those encountered during theorem \emph{application}. Suppose we have defined \lstinline|SemiGroup| as a structure predicate as follows\footnote{Note that defining \lstinline|SemiGroup| as a record instead of a straight conjunction does not make it any less of a predicate. The record form is simply more convenient in that it immediately gives us named projections for laws and substructures.}:
\begin{lstlisting}
  Record SemiGroup (G: Type) (e: relation G) (op: G → G → G): Prop :=
    { sg_setoid: Equivalence e
    ; sg_ass: Associative op
    ; sg_proper: Proper (e ==> e ==> e) op }.
\end{lstlisting}
Then by (1) making \lstinline|SemiGroup| a \emph{class} (by replacing the \lstinline|Record| keyword with the \lstinline|Class| keyword), (2) marking its proofs as \emph{instances} (by replacing the \lstinline|Lemma| keyword with the \lstinline|Instance| keyword), and (3) marking the \lstinline|SemiGroup| parameter of semigroup theorems as implicit (by using curly instead of round brackets), we no longer have to pass \lstinline|SemiGroup| proofs around manually ourselves, letting instance resolution do it for us instead. Because instance resolution is part of the unifier, this also works when the statement of the theorem we wish to apply only mentions some of the components (which admittedly doesn't make much sense for semigroups).

Next, we turn to problems concerning theorem \emph{declaration}. Our ideal will be the common mathematical vernacular, where one simply says:
\begin{quote}
Theorem: For $x, y, z$ in a semigroup $G$, $x * y * z = z * y * x$.
\end{quote}
(This silly statement allows us to clearly present the syntax.)

Without further support from the system, this would have to be written as
\begin{lstlisting}
  Theorem example G e op {P: SemiGroup G e op}:
    ∀ x y z, e (op (op x y) z) (op (op z y) x).
\end{lstlisting}
Because \lstinline|e| and \lstinline{op} are freshly introduced local names, we cannot bind notations to them prior to this theorem. Hence, if we want notations, what we really need are canonical names for these components. This is easily accomplished with single-field type classes containing one component each, which we will call \emph{operational type classes}\footnote{These single-field type classes are used in the same way in the \lstinline|Clean| standard library~\cite{clean}.}:
\begin{lstlisting}
  Class Equiv A := equiv: relation A.
  Class SemiGroupOp A := sg_op: A $\to$ A $\to$ A.

  Infix "=" := equiv: type_scope.
  Infix "&" := sg_op (at level 50, left associativity).
\end{lstlisting}
We use \lstinline|&| here and reserve the notation \lstinline|*| for (semi)ring multiplication.

As an aside, we note that the distinction between the class field name and the infix operator notation bound to it is really just a mildly awkward Coq artifact. In Haskell, where operators can themselves be used as names, there would be no need to have the \lstinline|equiv| and \lstinline|sg_op| names in addition to the operator `names'.

If we now retype \lstinline|SemiGroup| as:
\begin{lstlisting}
  ∀ (G: Type) (e: Equiv G) (op: SemiGroupOp G), Prop
\end{lstlisting}
then we can declare the theorem with:
\begin{lstlisting}
  Theorem example G e op {P: SemiGroup G e op}:
    ∀ x y z, x & y & z = z & y & x.
\end{lstlisting}
This works because instance resolution, invoked by the use of \lstinline|=| and \lstinline|&|, will find \lstinline|e| and \lstinline|op|, respectively. Hence, the above is really

\begin{lstlisting}
  Theorem example G e op {P: SemiGroup G e op}:
    ∀ x y z, equiv e (sg_op op (sg_op op x y) z) (sg_op op (sg_op op z y) x).
\end{lstlisting}
where \lstinline|e| and the \lstinline|op|s are filled in by instance resolution.

At this point, a legitimate worry might be that the \lstinline|Equiv|/\lstinline|SemiGroup| classes and their \lstinline|equiv|/\lstinline|sg_op| projections imply constant construction and deconstruction of records, harming the simplicity and flexibility of the predicate approach that we are trying so hard to preserve. No such construction and destruction takes place, however, because type classes with only a single field are not desugared into an actual record with field projections the way classes with any other number of fields are. Instead, both class itself and its field projection are defined as the identity function with a fancy type. Thus, the introduction of these canonical names is essentially free; the structure predicate's new type reduces straight back to what it was before.

A remaining eyesore in the theorem declaration is the enumeration of \lstinline|e| and \lstinline|op|. To remove these, we use a new parameter declaration feature called \emph{implicit generalization}, introduced in Coq specifically to support type classes. Using implicit generalization, we can write:
\begin{lstlisting}
  Theorem example `{SemiGroup G}: ∀ x y z: G, x & y & z = z & y & x.
\end{lstlisting}
The backtick tells Coq to insert implicit declarations of further parameters to \lstinline|SemiGroup G|, namely those declared as \lstinline|e| and \lstinline|op| above. It further lets us omit a name for the \lstinline|SemiGroup G| parameter itself as well. All of these will be given automatically generated names that we will never refer to.

Thus, we have reached the mathematical ideal we aimed for.

While we are on the topic of implicit generalization, we mention one inadequacy concerning their current implementation that we feel should be addressed for the facility to be a completely satisfying solution. While the syntax already supports variants (not shown above) that differ in how exactly different kinds of arguments are inferred and/or generalized, there is no support to have an argument be ``inferred if possible, generalized otherwise". The need for such a policy arises naturally when declaring a parameter of class type in a context where \emph{some} of its components are already available, while others are to be newly introduced. The current workaround in these cases involves providing names for components that are then never referred to, which is a bit awkward.

One aspect of the predicate approach we have not mentioned thus far is that in proofs parameterized by abstract structures, all components become hypotheses in the context. For the theorem above, the context looks like:
\begin{lstlisting}
  G: Type
  e: Equiv G
  op: SemiGroupOp G
  P: SemiGroup G e op
\end{lstlisting}
We are not particularly worried about overly large contexts, especially because most of the `extra' hypotheses we have compared to bundled approaches are declarations of relations, operators, and constants, which are all in some sense inert with respect to proof search. Hence, we do not foresee problems with large contexts for any but the most complex formalizations.

\subsection{Implicit syntax-directed algorithm composition}

Before we proceed with a discussion of the algebraic hierarchy based on predicate classes and operational classes, we briefly highlight one specific operational type class because we will use it later and because it is a particularly nice illustration of another neat application of operational type classes. The operation in question is that of deciding a proposition:
\begin{lstlisting}
  Class Decision (P: Prop): Type := decide: sumbool P (¬ P).
\end{lstlisting}
Here, \lstinline|sumbool| is just the (informative) sum of proofs.

\lstinline|Decision| is a very general-purpose type class that also works for predicates. For instance, to declare a parameter expressing decidability of, say, (setoid) equality on a type \lstinline|X|, we write: \lstinline|`{∀ a b: X, Decision (a = b)}|. To then use this (unnamed) decider to decide a particular equality, we simply say \lstinline|decide (x = y)|, and instance resolution will resolve the decider we declared.

With \lstinline|Decision| as a type class, we can very easily define composite deciders for things like conjunctions and quantifications over (finite) domains:
\begin{lstlisting}
  Instance decide_conj `{Decision P} `{Decision Q}: Decision (P ∧$$ Q).
\end{lstlisting}
With these in place, we can just say \lstinline|decide (x = y ∧$ $ p = q)| and let instance resolution automatically compose a decision procedure that can decide the specified proposition. This style of syntax-directed implicit composition of algorithms is very convenient and highly expressive.

\section{The algebraic hierarchy}\label{classes}

We have developed an algebraic hierarchy composed entirely out of predicate classes and operational classes as described in the previous section. For instance, our semiring interface looks as follows:
\begin{lstlisting}
  Class SemiRing A {e: Equiv A}
      {plus: RingPlus A} {mult: RingMult A}
      {zero: RingZero A} {one: RingOne A}: Prop :=
    { semiring_mult_monoid:> CommutativeMonoid A (op:=mult)(unit:=one)
    ; semiring_plus_monoid:> CommutativeMonoid A (op:=plus)(unit:=zero)
    ; semiring_distr:> Distribute mult plus
    ; semiring_left_absorb:> LeftAbsorb mult zero }.
\end{lstlisting}

All of \lstinline|Equiv|, \lstinline|RingPlus|, \lstinline|RingMult|, \lstinline|RingZero|, and \lstinline|RingOne| are operational (single-field) classes, with bound notations \lstinline|=|, \lstinline|+|, \lstinline|*|, \lstinline|0|, and \lstinline|1|, respectively.
\begin{wrapfigure}{r}{.4\textwidth}
    \includegraphics[scale=.8]{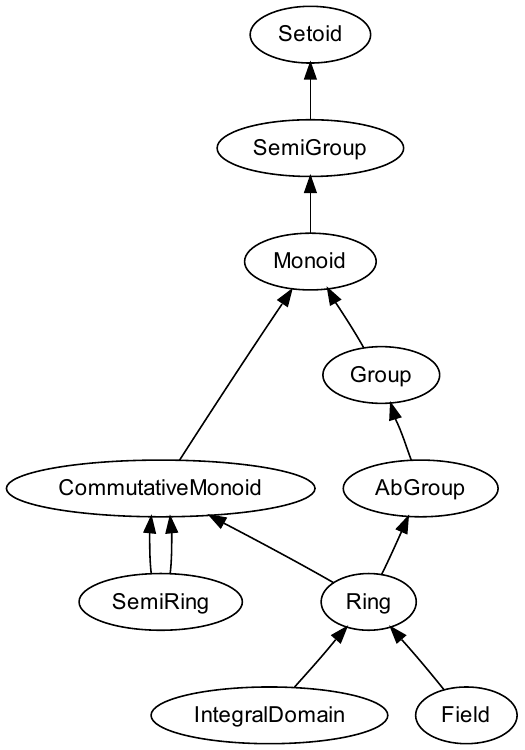}
\end{wrapfigure}
Let us briefly highlight some additional aspects of this style of structure definition in more detail.

Fields declared with \lstinline|:>| are registered as hints for instance resolution, so that in any context where \lstinline|(A, =, +, 0, *, 1)| is known to be a \lstinline|SemiRing|, \lstinline|(A, =, +, 0)| and \lstinline|(A, =, *, 1)| are automatically known to be \lstinline|CommutativeMonoid|s (and so on, transitively, because instance resolution is recursive). In our hierarchy, these substructures by themselves establish the inheritance diagram.

However, we can easily add additional inheritance relations by declaring corresponding class instances. For instance, while our \lstinline|Ring| class does not have a \lstinline|SemiRing| field, the following instance declaration has the exact same effect for the purposes of instance resolution (at least once proved, which is trivial):
\begin{lstlisting}
  Instance ring_as_semiring `{Ring R}: SemiRing R.
\end{lstlisting}

Thus, axiomatic structural properties and inheritance have precisely the same status as separately proved structural properties and inheritance, reflecting natural mathematical ideology. Again, contrast this with bundled approaches, where axiomatic inheritance relations determine projection paths, and where additional inheritance relations require rebundling and lead to additional and ambiguous projection paths for the same operations.

The declarations of the two inherited \lstinline|CommutativeMonoid| structures in \lstinline|SemiRing| nicely illustrate how predicate classes naturally support not just multiple inheritance, but \emph{overlapping} multiple inheritance, where the inherited structures may share components (in this case carrier and equivalence relation). The carrier \lstinline|A|, being an explicit argument, is specified as normal. The equivalence relation, being an implicit argument of class type, is resolved automatically to \lstinline|e|. The binary operation and constant would normally be automatically resolved as well, but we override the inference mechanism locally using Coq's existing named argument facility (which is only syntactic sugar of the most superficial kind) in order to explicitly pair multiplication with 1 for the first \lstinline|CommutativeMonoid| substructure, and addition with 0 for the second \lstinline|CommutativeMonoid| substructure. Again, contrast this with type system extensions such as Matita's manifest records, which are required to make this work when the records bundle components such as \lstinline|op| and \lstinline{unit} as \emph{fields} instead of parameters.

Since \lstinline|CommutativeMonoid| indirectly includes a \lstinline|SemiGroup| field, which in turn includes a \lstinline|Equivalence| field, having a \lstinline|SemiRing| proof means having two distinct proofs that the equality relation is an equivalence. This kind of redundant knowledge (which arises naturally) is never a problem in our setup, because the use of operational type classes ensures that terms composed of algebraic operations and relations never refer to structure proofs. We find that this is a tremendous relief compared to approaches that \emph{do} intermix the two and where one must be careful to ensure that such terms refer to the \emph{right} proofs of properties. There, even \emph{strong} proof irrelevance (which would make terms convertible that differ only in what proofs they refer to) would not make these difficulties go away entirely, because high-level tactics that rely on quotation of terms require syntactic identity (rather than `mere' convertibility) to recognize identical subterms.

Because predicate classes only provide contextual information and are insulated from the actual algebraic expressions, their instances can always be kept entirely opaque---only their existence matters. Together, these properties largely defuse an argument occasionally voiced against type classes concerning perceived unpredictability of instance resolution. While it is certainly true that in contexts with redundant information it can become hard to predict which instance of a predicate class will be found by proof search, it simply \emph{does not matter} which one is found. Moreover, for operational type classes the issue rarely arises because their instances are not nearly as abundant, and are systematically shared.

We use names for properties like distributivity and absorption, because these are type classes as well (which is why we declare their instances with \lstinline|:>|). It has been our experience that almost any generic predicate worth naming is worth representing as a predicate type class, so that its proofs will be resolved as instances behind the scenes whenever possible. Doing this consistently minimizes administrative noise in the code, bringing us closer to ordinary mathematical vernacular. Indeed, we believe that type classes provide an elegant and apt formalization of the seemingly casual manner in which ordinary mathematical presentation assumes implicit administration and use of a `database' of properties previously proved.

The operational type classes used in \lstinline|SemiRing| for zero, one, multiplication and addition, are the same ones used by \lstinline|Ring| and \lstinline|Field| (not shown). Thus, the realization that a particular semiring is in fact a ring or field has no bearing on how one refers to the operations in question, which is as it should be. The realization that a particular semigroup is part of a semiring \emph{does} call for a new (canonical) name, simply because of the need for disambiguation. The introduction of these additional names for the same operation is quite harmless in practice, because canonical names established by operational type class fields are identity functions, so that in most contexts the distinction reduces away instantly.

The hierarchy of predicate classes for the abstract structures themselves is mirrored by a hierarchy of predicate classes for morphisms. For instance:
\begin{lstlisting}
  Context `{Monoid A} `{Monoid B}.

  Class Monoid_Morphism (f: A → B) :=
    { monmor_from: Monoid A
    ; monmor_to: Monoid B
    ; monmor_sgmor:> SemiGroup_Morphism f
    ; preserves_mon_unit: f mon_unit = mon_unit }.
\end{lstlisting}

Some clarification is in order to explain the role of the \lstinline|Context| declaration of the two monoids. While \lstinline|Monoid_Morphism| seemingly depends on monoid-ness proofs (which would be a gross violation of our idiom), in fact it is only parameterized on the monoid \emph{components} declared through implicit generalization of the \lstinline|Monoid| declarations, because it only refers to those. Here, we use declarations of predicate class parameters merely as convenient shorthands to declare their components.

Notice that \lstinline|f| is \emph{not} made into an operational type class. The reason for this is that the role of \lstinline|f| is analogous to the carrier type in the previous predicate class definitions, in that it serves as the primary identification for the structure, and should therefore not be inferred.

We include the \lstinline|monmor_to| and \lstinline|monmor_from| fields because it does not make much sense to talk about monoid morphisms between non-monoids, and having these fields removes the need for \lstinline|Monoid| class constraints when we are already parameterizing definitions or theory on a \lstinline|Monoid_Morphism|. On the other hand, we will also wish to talk about monoid morphisms between \emph{known} monoids, and in these cases the fields will be strictly redundant. As mentioned before, it is a strength of our approach that such redundant knowledge is entirely harmless, so that we may freely posit these structural properties whenever they make sense and provide convenience, without risking rebundling tar-pits or projection path ambiguities down the line.

Unfortunately, there is actually one annoying wrinkle here, which will also explain why we do not register these two fields as instance resolution hints (by declaring them with \lstinline|:>|). What we really want these fields to express is ``\emph{if} in a certain context we know something to be a \lstinline|Monoid_Morphism|, \emph{then} realize that the source and target are \lstinline|Monoid|s''. However, the current instance resolution implementation has little support for this style of \emph{forward} reasoning, and is really primarily oriented on \emph{backward} reasoning: had we registered \lstinline|monmor_to| and \lstinline|monmor_from| as instance resolution hints, we would in fact be saying ``\emph{if} trying to establish that something is a \lstinline|Monoid|, \emph{then} try finding a \lstinline|Monoid_Morphism| to or from it'', which quickly degenerates into a wild goose chase. We will return to this point in section~\ref{conclusions}.

Having described the basic principles of our approach, in the remainder of this paper we tour other parts of our development, further illustrating what a state of the art formal development of foundational mathematical structures can look like with a modern proof assistant based on type theory.

These parts were originally motivated by our desire to cleanly express interfaces for basic numeric data types such as $\N$ and $\Z$ in terms of their categorical characterization as initial objects in the categories of semirings and rings, respectively. Let us start, therefore, with basic category theory.

\section{Category theory}\label{cat}

Following our idiom, we introduce operational type classes for the \emph{components} of a category:
\begin{lstlisting}
  Class Arrows (O: Type): Type := Arrow: O → O → Type.
  Class CatId O `{Arrows O} := cat_id: `(x ⟶ x).
  Class CatComp O `{Arrows O} :=
    comp: ∀ {x y z}, (y ⟶ z) → (x ⟶ y) → (x ⟶ z).

  Infix "⟶" := Arrow (at level 90, right associativity).
  Infix "◎" := comp (at level 40, left associativity).
\end{lstlisting}
(The categorical arrow is distinguished from the primitive function space arrow by its length.)

With these in place, our type class for categories follows the usual type-theoretical definition of a
category~\cite{saibi1995constructive}:
\begin{lstlisting}
  Class Category (O: Type) `{Arrows O} `{∀$$x y: O, Equiv (x ⟶ y)}
      `{CatId O} `{CatComp O}: Prop :=
    { arrow_equiv:> ∀ x y, Setoid (x ⟶ y)
    ; comp_proper:> ∀ x y z, Proper (equiv ==> equiv ==> equiv) comp
    ; comp_assoc w x y z (a: w ⟶ x) (b: x ⟶ y) (c: y ⟶ z):
        c ◎$$ (b ◎$$ a) = (c ◎$$ b) ◎$$ a
    ; id_l `(a: x ⟶ y): cat_id ◎$$ a = a
    ; id_r `(a: x ⟶ y): a ◎$$ cat_id = a }.
\end{lstlisting}
This definition is based on the 2-categorical idea of having equality only on arrows, not on objects.

Initiality, too, is defined by a combination of an operational and a predicate class:
\begin{lstlisting}
  Context `{Category X}.
  Class InitialArrows (x: X): Type := initial_arrow: ∀ y, x ⟶ y.
  Class Initial (x: X) `{InitialArrows x}: Prop :=
    initial_arrow_unique: ∀ y (a: x ⟶ y), a = initial_arrow y.
\end{lstlisting}
The operational class \lstinline|InitialArrows| designates the arrows that originate from an initial object \lstinline|x| by virtue of it being initial. The \lstinline|Initial| class itself further requires these ``initial arrows'' to be unique. Having \lstinline|InitialArrows| as an operational type class means that we can always simply say \lstinline|initial_arrow y| whenever \lstinline|y| is known to be an object in a category known to have an initial object (where `known' should be read as ``can be determined by instance resolution'').

Strictly speaking the above is all we need in order to continue with the story line leading up to the numerical interfaces, but just to give a further taste of what category theory with this setup looks like in practice, we briefly mention a few more definitions and theorems.

\subsection{Functors}

In our definition of functors we see the by now familiar refrain once more:
\begin{lstlisting}
  Context `{Category C} `{Category D} (map_obj: C → D).

  Class Fmap: Type :=
    fmap: ∀ {v w: C}, (v ⟶ w) → (map_obj v ⟶ map_obj w).

  Class Functor `{Fmap}: Prop :=
    { functor_from: Category C
    ; functor_to: Category D
    ; functor_morphism:> ∀ a b: C, Setoid_Morphism (@fmap _ a b)
    ; preserves_id: `(fmap (cat_id: a ⟶ a) = cat_id)
    ; preserves_comp `(f: y ⟶ z) `(g: x ⟶ y):
      fmap (f ◎$$ g) = fmap f ◎$$ fmap g }.
\end{lstlisting}
We ought to say a few words about our use of \lstinline|fmap|. The usual mathematical notational convention for functor application is to use the name of the functor to refer to both its object map and its arrow map, relying on additional conventions regarding object/arrow names for disambiguation: \lstinline|F x| and \lstinline|F f| map an object and an arrow, respectively, because \lstinline|x| and \lstinline|f| are conventional names for objects and arrows, respectively.

In Coq, for a term \lstinline|F| to function as though it had two different types simultaneously (namely the object map and the arrow map), there must either (1) be coercions from the type of F to either function, or (2) F must be (coercible to) a single function that is able to consume both object and arrow arguments. In addition to not being supported by Coq, option (1) would violate our policy of leaving components unbundled. For (2), if it could be made to work at all, F would need a pretty egregious type considering that arrow types are indexed by objects, and that the type of the arrow map
\begin{lstlisting}
 ∀ x y, (x ⟶ y) → (F x ⟶ F y)
\end{lstlisting}
must refer to the object map.

We feel that these issues are not limitations of the Coq system, but merely reflect the fact that notationally identifying these two distinct and interdependent maps is an abuse of notation of sufficient severity to make it ill-suited to a formal development where software engineering concerns apply. Hence, we do not adopt this practice, and use \lstinline|fmap F| (name taken from the Haskell standard library) to refer to the arrow map of a functor \lstinline|F|.

\subsection{Natural transformations and adjunctions}

We introduce a convenient notation for the type of the computational content of a natural transformation between two functors:
\begin{lstlisting}
  Notation "F ⇛ G" := (∀ x, F x ⟶ G x).
\end{lstlisting}
Now assume the following context:
\begin{lstlisting}
  Context `{Category C} `{Category D}
    `{Functor (F: C → D)} `{Functor (G: D → C)}.
\end{lstlisting}
The naturality property is easy to write:
\begin{lstlisting}
  Class NaturalTransformation (η: F ⇛ G): Prop :=
    { naturaltrans_from: Functor F
    ; naturaltrans_to: Functor G
    ; natural: ∀ `(f: x ⟶ y), η$$ y ◎$$ fmap F f = fmap G f ◎$$ η$$ x }.
\end{lstlisting}

Adjunctions can be defined in different ways. A nice symmetric definition is the following:
\begin{lstlisting}
  Class Adjunction (φ: ∀ `(F c ⟶ d), (c ⟶ G d)): Prop :=
    { adjunction_left_functor: Functor F _
    ; adjunction_right_functor: Functor G _
    ; natural_left `(f: d ⟶ d') c: (fmap G f ◎) ∘ φ$$ = φ (c:=c) ∘ (f ◎)
    ; natural_right `(f: c' ⟶ c) d: (◎ f) ∘ φ (d:=d) = φ$$ ∘ (◎ fmap F f) }.
\end{lstlisting}
An alternative definition is the following:
\begin{lstlisting}
  Class AltAdjunction (η: id ⇛ G ∘ F) (φ: ∀ `(f: c ⟶ G d), F c ⟶ d): Prop :=
    { alt_adjunction_natural_unit: NaturalTransformation η
    ; alt_adjunction_factor: ∀ `(f: c ⟶ G d),
        is_sole ((f =) ∘ (◎ η$$ c) ∘ fmap G) (φ f) }.
\end{lstlisting}
Formalizing the (nontrivial) proof that these two definitions are equivalent provides a nice test for our definitions. As a first step, we have constructed the unit and co-unit of the adjunction, thus proving MacLane's Theorem 1. We have concisely and closely followed his proof~\cite{CatWork}. 

\section{Universal algebra}\label{univ}

To specify the natural numbers and the integers as initial objects in the categories of semirings and rings, respectively, definitions of these categories are needed. While one could define both of them manually, greater economy can be achieved by recognizing that both semirings and rings can be defined by equational theories, for which \emph{varieties} can be defined generically. Varieties are categories consisting of models for a fixed theory with homomorphisms between them.

To this end, we have formalized some of the theory of multisorted universal algebra and equational theories. We chose not to revive existing formalizations~\cite{DBLP:conf/tphol/Capretta99,dominguez2008formalizing} of universal algebra, because an important aim for us has been to find out what level of elegance, convenience, and integration can be achieved by leveraging the state of the art in Coq facilities (of which type classes are the most important example).

\subsection{Signatures and algebras}

A multisorted signature enumerates sorts, operations, and specifies the `types' of the operations as non-empty lists of sorts, where the final element denotes the result type:
\begin{lstlisting}
  Inductive Signature: Type :=
    { sorts: Set
    ; operation:> Set
    ; operation_type:> operation → ne_list sorts }.
\end{lstlisting}
Given an interpretation of the sorts (mapping each symbolic sort to a carrier type), interpretations of the operations are easily represented by an operational type class:
\begin{lstlisting}
  Variables (σ: Signature) (carriers: sorts σ$$ → Type).

  Class AlgebraOps :=
    algebra_op: ∀ o: operation σ, fold (→$\hspace{-1mm}$) (map carriers (operation_type σ$$ o)).
\end{lstlisting}
Because our carriers will normally be equipped with a setoid equality, we further define the predicate class \lstinline|Algebra|, stating that each of the operations respects the setoid equality on the carriers:
\begin{lstlisting}
  Class Algebra `{∀$$a, Equiv (carriers a)} `{AlgebraOps}: Prop :=
      { algebra_setoids:> ∀ a, Setoid (carriers a)
      ; algebra_propers:> ∀ o: σ, Proper (=) (algebra_op o) }.
\end{lstlisting}
The \lstinline|(=)| referred to in \lstinline|algebra_propers| is an automatically derived \lstinline|Equiv| instance expressing setoid-respecting extensionality for the function types produced by the \lstinline|fold| in \lstinline|AlgebraOps|.

We do not unbundle \lstinline|Signature| because it represents a triple that will always be specifically constructed for subsequent use with the universal algebra facilities. We have no ambition to recognize signature triples ``in the wild'', nor will we ever talk about multiple signatures sharing sort- or operation enumerations.

\subsection{Equational theories and varieties}
\label{varieties}

To adequately characterize structures such as semirings and rings, we need not just a signature that enumerates and gives the types of their operations, but also a specification of the axioms (laws) that these operations must satisfy. For this, we define \lstinline|EquationalTheory| as a signature together with a set of laws, the latter represented by a predicate over equality entailments:
\begin{lstlisting}
  Record EquationalTheory :=
    { eqt_sig:> Signature
    ; eqt_laws:> EqEntailment eqt_sig → Prop }.
\end{lstlisting}
An \lstinline|EqEntailment| consists of premises and a conclusion represented by an inductively defined statement grammar, which in turn uses an inductively defined term grammar. A detailed discussion of these definitions and the theory developed for them is beyond the scope of this paper.

We now introduce a predicate class designating algebras that satisfy the laws of an equational theory:
\begin{lstlisting}
  Class InVariety
    (et: EquationalTheory) (carriers: sorts et → Type)
    {e: ∀ a, Equiv (carriers a)} `{AlgebraOps et carriers}: Prop :=
    { variety_algebra:> Algebra et carriers
    ; variety_laws: ∀ s, eqt_laws et s → (∀$$vars, eval_stmt et vars s) }.
\end{lstlisting}

What remains is to show that carrier sets together with \lstinline|Equiv|s and \lstinline|AlgebraOps| satisfying \lstinline|InVariety| for a given \lstinline|EquationalTheory| do indeed form a \lstinline|Category| (the `variety'). Since we need a type for the objects in the \lstinline|Category|, at this point we have no choice but to bundle components and proof together in a record:
\begin{lstlisting}
  Variable et: EquationalTheory.

  Record ObjectInVariety: Type := object_in_variety
    { variety_carriers:> sorts et → Type
    ; variety_equiv: ∀ a, Equiv (variety_carriers a)
    ; variety_op: AlgebraOps et variety_carriers
    ; variety_proof: InVariety et variety_carriers }.
\end{lstlisting}
The arrows will be homomorphisms, which are also defined generically for any equational theory:

\begin{lstlisting}
  Instance: Arrows Object := λ X Y: Object => sig (HomoMorphism et X Y).
\end{lstlisting}

The instance definitions for identity arrows, arrow composition, arrow setoid equality, and composition propriety, are all trivial, as is the final \lstinline|Category| instance:
\begin{lstlisting}
  Instance: Category ObjectInVariety.
\end{lstlisting}

In addition to this variety category, we also have categories of lawless algebras, as well as forgetful functors from the former to the latter, and from the latter to the category of setoids.

\subsection{The first homomorphism theorem}
\label{homothm}

To give a further taste of what universal algebra in our development looks like, we consider the definitions involved in the first homomorphism theorem~\cite{meinke1993universal} in more detail:
\begin{theorem}[First homomorphism theorem]
If $A$ and $B$ are algebras, and $f$ is a homomorphism from $A$ to $B$, then the equivalence relation $\sim$ defined by ``$a\sim b \leftrightarrow f(a)=f(b)$'' is a congruence on $A$, and the quotient algebra $A/\hspace{-1mm}\sim$ is isomorphic to the image of $f$, which is a subalgebra of B.
\end{theorem}

A set of relations \lstinline|e| (one for each sort) is a congruence for an existing algebra if (1) \lstinline|e| respects that algebra's existing setoid equality, and (2) the operations with \lstinline|e| again form an algebra (namely the quotient algebra):
\begin{lstlisting}
  Context `{Algebra σ$$ A}.
  Class Congruence (e: ∀ s: sorts σ, relation (v s)): Prop :=
    { congruence_proper:> ∀ s, Proper (equiv ==> equiv ==> iff) (e s)
    ; congruence_quotient:> Algebra σ$$ v (e:=e) }.
\end{lstlisting}
We have proved that this natural and economical type-theoretic formulation that leverages our systematic integration of setoid equality is equivalent to the traditional definition of congruences as relations that, represented as sets of pairs, form a subalgebra of the product algebra.

For the homomorphism theorem, we begin by declaring our dramatis personae:
\begin{lstlisting}
  Context `{HomoMorphism σ$$ A B f}.
\end{lstlisting}
With \lstinline|~| defined as indicated, the first part of the proof is simply the definition of the following instance:
\begin{lstlisting}
  Instance co: Congruence σ$$ (~).
\end{lstlisting}

For the second part, we describe the image of \lstinline|f| as a predicate over \lstinline|B|, and show that it is closed under the operations of the algebra: 
\begin{lstlisting}
  Definition image s (b: B s): Type := sigT (λa => f s a = b).

  Instance: ClosedSubset image.
\end{lstlisting}
The \lstinline|sigT| type constructor is a \lstinline|Type|-sorted existential quantifier. \lstinline|ClosedSubset| is defined elsewhere as:
\begin{lstlisting}
  Context `{Algebra σ$$ A} (P: ∀ s, A s → Type).
  Class ClosedSubset: Type :=
    { subset_proper: ∀ s x x', x = x' → iffT (P s x) (P s x')
    ; subset_closed: ∀ o, op_closed (algebra_op o) }.
\end{lstlisting}
Here, \lstinline|op_closed| is defined by recursion over the symbolic operation types.

The reason we define \lstinline|image| and \lstinline|ClosedSubset| in \lstinline|Type| rather than in \lstinline|Prop| is that since the final goal of the proof is to establish an isomorphism in the category of σ-algebras (where arrows are algebra homomorphisms), we will eventually need to map elements in the subalgebra defined by \lstinline|image| back to their pre-image in \lstinline|A|.

However, there are contexts (in other proofs) where \lstinline|Prop|-sorted construction of subalgebras really \emph{is} appropriate. Unfortunately, Coq's universe polymorphism is not yet up to the task of letting us use a single set of definitions to handle both cases. In particular, there is no universe polymorphism for ordinary definitions (as opposed to inductive definitions) yet. We will return to this point later. In our development, we have two sets of definitions, one for \lstinline|Prop| and one for \lstinline|Type|, resulting in duplication of about a hundred lines of code.

For the main theorem, we now bundle the quotient algebra and the subalgebra into records akin to \lstinline|ObjectInVariety| from section~\ref{varieties}:
\begin{lstlisting}
  Definition quot_obj: algebra.Object σ :=
    algebra.object σ A (algebra_equiv:=(~)).
  Definition subobject: algebra.Object σ$$ :=
    algebra.object σ (ua_subalgebraT.carrier image).
\end{lstlisting}
Here, \lstinline|algebra| is the module defining the bundled algebra record \lstinline|Object| with constructor \lstinline|object|. The module \lstinline|ua_subalgebraT| constructs subalgebras.

Finally, we define a pair of arrows between the two and show that these arrows form an isomorphism:
\begin{lstlisting}
   Program Definition back: subobject ⟶ quot_obj
    := λ _ X => projT1 (projT2 X).

  Program Definition forth: quot_obj ⟶ subobject
    := λ a X => existT _ (f a X) (existT _ X (reflexivity _)).

  Theorem first_iso: iso_arrows back forth.
\end{lstlisting}
The \lstinline|Program| command generates proof obligations (not shown) expressing that these two arrows are indeed homomorphisms. The proof of the theorem itself is trivial.

\section{Numerical interfaces}\label{numbers}

\lstinline|EquationalTheory|'s for semirings and rings are easy to define, and so from section~\ref{varieties} we get corresponding categories in which we can postulate initial objects:
\begin{lstlisting}
  Class Naturals (A: ObjectInVariety semiring_theory) `{InitialArrow A}: Prop :=
    { naturals_initial:> Initial A }.
\end{lstlisting}
While succinct, this definition is not a satisfactory abstraction because the use of \lstinline|ObjectInVariety| for the type of the \lstinline|A| component `leaks' the fact that we used this one particular universal algebraic construction of the category, which is just an implementation choice. Furthermore, this definition needs an additional layer of class instances to relate it to the \lstinline|SemiRing| class from our algebraic hierarchy.

What we \emph{really} want to say is that an implementation of the natural numbers ought to be an a-priori \lstinline|SemiRing| that, when \emph{bundled} into an \lstinline|ObjectInVariety semiring_theory|, is initial in said category. This is a typical example where conversion functions between concrete classes such as \lstinline|SemiRing| and instantiations of more abstract classes such as \lstinline|InVariety| and \lstinline|Category| are required in our development in order to leverage and apply concepts and theory defined for the latter to the former. While sometimes a source of some tension in that these conversions are not yet applied completely transparently whenever needed, the ability to move between ``down to earth'' and ``high in the sky'' perspectives on the same abstract structures has proved invaluable in our development, and we will give more examples of this in a moment.

Taking these conversion functions for granted, we will also need a ``down to earth'' representation of the initiality arrows if we are to give a \lstinline|SemiRing|-based definition of the interface for natural numbers. Once again, we introduce an operational type class to represent this particular component:
\begin{lstlisting}
  Class NaturalsToSemiRing (A: Type) :=
    naturals_to_semiring: ∀ B `{RingMult B} `{RingPlus B} `{RingOne B}
      `{RingZero B}, A → B.
\end{lstlisting}
The instance for \lstinline|nat| is defined as follows:
\begin{lstlisting}
Instance nat_to_semiring: NaturalsToSemiRing nat :=
  λ _ _ _ _ _ => fix f (n: nat) := match n with 0 => 0 | S m => f m + 1 end.
\end{lstlisting}

To use \lstinline|NaturalsToSemiRing| with \lstinline|Initial|, we define an additional conversion instance that takes a \lstinline|NaturalsToSemiRing| along with a proof showing that it yields \mbox{\lstinline|SemiRing_Morphism|s}, and builds an \lstinline|InitialArrow| instance out of it. This conversion instance in turn invokes another conversion function that translates concrete \lstinline|SemiRing_Morphism| proofs into univeral algebra \lstinline|Homomorphism|s instantiated with the semiring signature, which make up the arrows in the category.

With these instances in place, we can now define the improved natural numbers specification:
\begin{lstlisting}
  Context `{SemiRing A} `{NaturalsToSemiRing A}.
  Class Naturals: Prop :=
    { naturals_ring:> SemiRing A
    ; naturals_to_semiring_mor:> ∀ `{SemiRing B},
        SemiRing_Morphism (naturals_to_semiring A B)
    ; naturals_initial:> Initial (bundle_semiring A) }.
\end{lstlisting}
Basing theory and programs on this abstract interface instead of on specific implementation (such as the ubiquitous Peano naturals \lstinline|nat| in the Coq standard library) is not only cleaner mathematically, but also facilitates easy swapping between implementations. And this benefit is far from theoretical, as diverse representations of the natural numbers are abound; for instance, unary, binary, factor multisets, and arrays of native machine words.

Since initial objects in categories are isomorphic, we can easily derive that \lstinline|naturals_to_semiring| gives isomorphisms between different \lstinline|Naturals| implementations:
\begin{lstlisting}
  Lemma iso_naturals `{Naturals A} `{Naturals B}:
    ∀ a: A, naturals_to_semiring B A (naturals_to_semiring A B a) = a.
\end{lstlisting}
This is very useful, because some properties of and operations on naturals are more easily proved, respectively defined, for concrete implementations (such as \lstinline|nat|) and then \emph{lifted} to the abstract \lstinline|Naturals| interface so that they work for arbitrary implementations. For example, while showing decidability for an arbitrary \lstinline|Naturals| implementation directly is tricky, it is very easy to show decidability for \lstinline|nat|. Using \lstinline|iso_naturals|, the latter can be very straightforwardly used to implement the former.

To lift properties such as injectivity of partially applied addition and multiplication from \lstinline|nat| to arbitrary \lstinline|Naturals| implementations, we take a longer detour. As part of our universal algebra theory, we have proved that proofs of statements in the language of an equational theory can be transferred between isomorphic implementations. Hence, we can transfer proofs of such statements between implementations of \lstinline|Naturals|, requiring only that we reflect the concrete statement (expressed in terms of the operational type classes) to a symbolic statement in the language of semirings. We intend to eventually make this reflection completely automatic using type class based quotation techniques along the lines of those described in Section~\ref{quoting}.

Thanks to our close integration of universal algebra we can actually obtain a \lstinline|Naturals| implementation completely automatically, by invoking a generic construction of initial models built from the closed term algebra for the signature along with a setoid equality expressing the congruence closure of the identities in the equational theory. However, this implementation is not very useful, neither in terms of efficiency, nor as a canonical implementation (to be used as the basis for theory and programs that are then subsequently lifted). For example, defining a normalization procedure to decide the aforementioned setoid equality is far harder than deciding equality for, say, \lstinline|nat|.

\subsection{Specialization}

The generic \lstinline|Decision| instance for \lstinline|Naturals| equality implemented by mapping to \lstinline|nat| will typically be far less efficient than a specialized implementation for a particular representation of the natural numbers. Fortunately, with Coq's type classes it is no problem for instances overlapping in this way to co-exist. We can even deprioritize the generic instance so that instance resolution will always pick the specialization when the representation is known.

To permit a \emph{generic} function operating on naturals to take advantage of specialized operations, we simply introduce an additional instance parameter:
\begin{lstlisting}
Definition calculate_things `{Naturals N} `{∀$$n m: N, Decision (n = m)}
  (a b: nat): ... := ... decide (a = b) ... .
\end{lstlisting}
Without the \lstinline|Decision| parameter \lstinline|calculate_things| would be equally correct, but could be less efficient. Thus, by this scheme one can start by writing correct-but-possibly-inefficient programs that make use of generic operation instances, and then selectively improve efficiency of key algorithms simply by adding additional operational type class instance parameters where profiling shows it to make a significant difference, without changing their definition body.

Other examples of operations on natural numbers that are sensible choices for specialization include: subtraction, distance, and division and multiplication by 2.

\subsection{Integers, rationals, and polynomials}

The abstract interface for integers is completely analogous to the one for natural numbers:
\begin{lstlisting}
  Context `{Ring A} `{IntegersToRing A}.
  Class Integers: Prop :=
    { integers_ring:> Ring A
    ; integers_to_ring_mor:> ∀ `{Ring B},
        Ring_Morphism (integers_to_ring A B)
    ; integers_initial:> Initial (ring.object A) }.
\end{lstlisting}

The rationals are characterized as a decidable field with an injective ring morphism from a canonical implementation of the integers and a surjection of fractions of such integers:
\begin{lstlisting}
  Context `{Field A} `{∀$$x y: A, Decision (x = y)} {inj_inv}.
  Class Rationals: Prop :=
    { rationals_field:> Field A
    ; rationals_frac: Surjective
        (λ p => integers_to_ring (Z nat) A (fst p) *
          / integers_to_ring (Z nat) A (snd p)) (inv:=inj_inv)
    ; rationals_embed_ints: Injective (integers_to_ring (Z nat) A) }.
\end{lstlisting}
Here, \lstinline|Z| is an \lstinline|Integers| implementation paramerized by a \lstinline|Naturals| implementation, for which we just take \lstinline|nat|. The choice of \lstinline|Z nat| here is immaterial; we could have picked another, or even a generic, implementation of \lstinline|Integers|, but doing so would provide no benefit.

In our development we prove that the standard library's default rationals do indeed implement \lstinline|Rationals|, as do implementations of the \lstinline|QType| module interface. While the latter is rather ad-hoc from a theoretical perspective, it is nevertheless of great practical interest because it is used for the very efficient \lstinline|BigQ| rationals based on machine integers~\cite{machineintegers}. Hence, theory and programs developed on our \lstinline|Rationals| interface applies and can make immediate use of these efficient rationals. We plan to rebase the computable real number implementation~\cite{Oconnor:real} on this interface, precisely so that it may be instantiated with efficient implementations like these.

We also plan to provide an abstract interface for polynomials as a free commutative algebra. This would unify existing implementations such as coefficient lists and Bernstein polynomials; see~\cite{ZumkellerPhD} for the latter.

\section{Quoting with type classes}\label{quoting}

A common need when interfacing generic theory and utilities developed for algebraic structures (such as normalization procedures) with concrete instances of these structures is to take a concrete expression or statement in a model of a particular algebraic structure, and translate it to a symbolic expression or statement in the language of the algebra's signature, so that its structure can be inspected.

Traditionally, proof assistants such as Coq have provided sophisticated tactics or built-in commands to support such \emph{quoting}. Unification hints~\cite{Hints}, a very general way of facilitating user-defined extensions to term and type inference, can be used to semi-automatically build quote functions without dropping to a meta-level.\footnote{Gonthier provides similar functionality by ingeniously using canonical structures.} This feature is absent from Coq, but fortunately type classes also allow us to do this, as we will now show.

For ease of presentation we show only a proof of concept for a very concrete language. We are currently working to integrate this technique with our existing universal algebra infrastructure. In particular, the latter's term data type should be ideally suited to serve as a generic symbolic representation of terms in a wide class of algebras. This should let us implement the basic setup of the technique once and for all, so that quotation for new algebraic structures can be enabled with minimal effort.

For the present example, we define an ad-hoc term language for monoids
\begin{lstlisting}
  Inductive Expr (V: Type) := Mult (a b: Expr V) | One | Var (v: V).
\end{lstlisting}
The expression type is parameterized over the set of variable indices. Below we use an implicitly defined heap of such variables. Hence, we diverge from~\cite{Hints}, which uses \lstinline|nat| for variable indices, thereby introducing a need for dummy variables for out-of-bounds indices.

Suppose now that we want to quote \lstinline|nat| expressions built from 1 and multiplication. To describe the relation we want the symbolic expression to have to the original expression, we first define how symbolic expressions evaluate to values (given a variable assignment):
\begin{lstlisting}
  Definition Value := nat.
  Definition Env V := V → Value.

  Fixpoint eval {V} (vs: Env V) (e: Expr V): Value :=
    match e with
    | One => 1
    | Mult a b => eval vs a * eval vs b
    | Var v => vs v
    end.
\end{lstlisting}

We can now state our goal: given an expression of type \lstinline|nat|, we seek to construct an \lstinline|Expr V| for some appropriate \lstinline|V| along with a variable assignment, such that evaluation of the latter yields the former. Because we will be doing this incrementally, we introduce a few simple variable ``heap combinators'':
\begin{lstlisting}
  Definition novars: Env False := False_rect _.
  Definition singlevar (x: Value): Env unit := λ _ => x.
  Definition merge {A B} (a: Env A) (b: Env B): Env (A+B) :=
    λ i => match i with inl j => a j | inr j => b j end.
\end{lstlisting}
These last two combinators are the `constructors' of an implicitly defined subset of Gallina terms, representing heaps, for which we will implement syntactic lookup with type classes in a moment. The heap can also be defined explicitly, with no essential change in the code.

With these, we can define the primary ingredient, the \lstinline|Quote| class:
\begin{lstlisting}
  Class Quote {V} (l: Env V) (n: Value) {V'} (r: Env V'): Type :=
    { quote: Expr (V + V')
    ; eval_quote: eval (merge l r) quote = n }.
\end{lstlisting}
We can think of \lstinline|Quote| as the type for a family of Prolog-like syntax-directed resolution functions, which will take as input \lstinline|V| and \lstinline|l| representing previously encountered holes (opaque subexpressions that could not be destructured further) and their values, along with a concrete term \lstinline|n| to be quoted. Their `output' will consist not only of the fields in the class, but also of \lstinline|V'| and \lstinline|r| representing additional holes and their values. Hence, a type class constraint of the form \lstinline|Quote x y z| should be read as ``quoting \lstinline|y| with existing heap \lstinline|x| generates new heap \lstinline|z|''.

The \lstinline|Quote| instance for 1 illustrates the basic idea:
\begin{lstlisting}
  Instance quote_one V (v: Env V): Quote v 1 novars := { quote := One }.
\end{lstlisting}
The expression `1' can be quoted in any context \lstinline|(V, v)|, introduces no new variables, and the symbolic term representing it is just \lstinline|One|. The \lstinline|eval_quote| field is turned into a trivial proof obligation.

The \lstinline|Quote| instance for multiplication is a little more subtle, but really only does a bit of heap juggling:
\begin{lstlisting}
  Instance quote_mult V (v: Env V) n V' (v': Env V') m V'' (v'': Env V'')
    `{Quote v n v'} `{Quote (merge v v') m v''}:
      Quote v (n * m) (merge v' v'') :=
      { quote :=
        Mult (map_var shift (quote n)) (map_var sum_assoc (quote m)) }.
\end{lstlisting}

These two instances specify how 1 and multiplications are to be quoted, but what about other expressions? For these, we want to distinguish two kinds: expressions we have seen before, and those we have not. To make this distinction, we need to be able to look up expressions in variable heaps to see if they're already there. Importantly, we must do this not by comparing the values they evaluate to, but by actually browsing the term denoting the variable heap --- that is, a composition from \lstinline|novars|, \lstinline|singlevar|, and \lstinline|merge|. This, too, is a job for a type class:
\begin{lstlisting}
  Class Lookup {A} (x: Value) (v: Env A) := { key: A; key_correct: v key = x }.
\end{lstlisting}
Our first \lstinline|Lookup| instance states that \lstinline|x| can be looked up in \lstinline|singlevar x|:
\begin{lstlisting}
  Instance singlevar_lookup (x: Value): Lookup x (singlevar x) := { key := tt }.
\end{lstlisting}
Finally, if an expression can be looked up in a pack, then it can also be looked up when that pack is merged with another pack:
\begin{lstlisting}
  Context (x: Value) {A B} (va: Env A) (vb: Env B).

  Instance lookup_left `{Lookup x va}: Lookup x (merge va vb)
    := { key := inl (key x va) }.

  Instance lookup_right `{Lookup x vb}: Lookup x (merge va vb)
    := { key := inr (key x vb) }.
\end{lstlisting}

With \lstinline|Lookup|, we can now define a \lstinline|Quote| instance for previously encountered expressions:
\begin{lstlisting}
  Instance quote_old_var V (v: Env V) x {Lookup x v}:
    Quote v x novars | 8 := { quote := Var (inl (key x v)) }.
\end{lstlisting}

If none of the \lstinline|Quote| instances defined so far apply, the term in question is a newly encountered hole. For this case we define a catch-all instance with a low priority, which yields a singleton heap containing the expression:
\begin{lstlisting}
  Instance quote_new_var V (v: Env V) x: Quote v x (singlevar x) | 9
    := { quote := Var (inr tt) }.
\end{lstlisting}
And with that, we can start quoting:
\begin{lstlisting}
  Goal ∀ x y (P: Value → Prop), P ((x * y) * (x * 1)).
    intros.
    rewrite <- eval_quote.
\end{lstlisting}
The \lstinline|rewrite| rewrites the goal to (something that reduces to):
\begin{lstlisting}
  P (eval
     (merge novars
        (merge (merge (singlevar x) (singlevar y)) (merge novars novars)))
     (Mult (Mult (Var (inr (inl (inl ())))) (Var (inr (inl (inr ())))))
        (Mult (Var (inr (inl (inl ())))) One)))
\end{lstlisting}

The following additional utility lemma lets us quote equalities with a shared heap (so that an opaque expression that occurs on both sides of the equation is not represented by two distinct variables):
\begin{lstlisting}
  Lemma quote_equality {V} {v: Env V} {V'} {v': Env V'} (l r: Value)
    `{Quote novars l v} `{Quote v r v'}:
      let heap := merge v v' in
      eval heap (map_var shift quote) = eval heap quote → l = r.
\end{lstlisting}

Notice that we have not made any use of \lstinline|Ltac|, Coq's tactic language. Instead, we have used instance resolution as a unification-based programming language to steer the unifier into inferring the symbolic quotation.

\section{Sequences and universes}\label{sequences}

Finite sequences are another example of a concept that can be represented in many different ways: as cons-lists, maps from bounded naturals, array-queues, etc. Here, too, the introduction of an abstract interface facilitates implementation independence.

Mathematically, finite sequences can be characterized as free monoids over sets. A categorical way of expressing this is in terms of adjunctions. As with the numeric interfaces, we \emph{could} fully embrace this perspective, paying no heed to practicality of implementation and usage, and define a relatively succinct type class for sequences as follows:
\begin{lstlisting}
  Class PoshSequence
    (free: setoid.Object → monoid.Object) `{Fmap free}
    (singleton: id ⇛ monoid.forget ∘ free)
    (extend: `((x ⟶ monoid.forget y) → (free x ⟶ y))): Prop :=
    { sequence_adjunction: AltAdjunction singleton extend
    ; extend_morphism: `(Setoid_Morphism (extend x y)) }.
\end{lstlisting}
Here, \lstinline|monoid.forget| is the forgetful functor from monoids to sets.

However, we \emph{do} care about practicality, and so we will again take a more concrete perspective, starting with operational type classes for the characteristic operations:
\begin{lstlisting}
  Context `{Functor (seq: Type → Type)}.

  Class Extend := extend: ∀ {x y} `{SemiGroupOp y} `{MonoidUnit y},
    (x → y) → (seq x → y).
  Class Singleton := singleton: ∀ x, x → seq x.
\end{lstlisting}
With these, we can define the predicate class for sequences:
\begin{lstlisting}
  Class Sequence
   `{∀ a, MonoidUnit (seq a)} `{∀ a, SemiGroupOp (seq a)}
   `{∀ a, Equiv a → Equiv (seq a)} `{Singleton} `{Extend}: Prop := ...
\end{lstlisting}
On top of this interface we can build theory about typical sequence operations such as maps, folds, their relation to \lstinline|singleton| and \lstinline|extend|, et cetera. We can also generically define `big operators' for sums ($\sum$) and products ($\prod$) of sequences, and easily show properties like distributivity, all without ever mentioning cons-lists.

Unfortunately, disaster strikes when, after having defined this theory, we try to show that regular cons-lists implement the abstract \lstinline|Sequence| interface. When we get to the point where we want to define the \lstinline|Singleton| operation, Coq emits a universe inconsistency error. The problem is that because of the categorical constructions involved, the theory forces \lstinline|Singleton| to inhabit a relatively high universe level, making it incompatible with lowly \lstinline|list|.

Universe polymorphism could in principle most likely solve this problem, but its current implementation in Coq only supports universe polymorphic \emph{inductive} definitions, while \lstinline|Singleton| is a regular definition. Universe polymorphic regular definitions have historically not been supported in Coq, primarily because of efficiency concerns. However, we have taken up the issue with the Coq development team, and they have agreed to introduce the means to let one turn on universe polymorphism for definitions voluntarily on a per-definition basis. With such functionality we could make \lstinline|Singleton| universe polymorphic, and hopefully resolve these problems.

In other places in our development, too, we have encountered universe inconsistencies that could be traced back to universe monomorphic definitions being forced into disparate universes (\lstinline|Equiv| being a typical example). Hence, we consider the support for universe polymorphic definitions that is currently being implemented to be of great importance to the general applicability and scalability of our approach.
 
\section{Conclusions}\label{conclusions}

While bundling operational and propositional components of abstract structures into records may seem natural at first, doing so actually introduces many serious problems. With type classes we avoid these problems by avoiding bundling altogether.

It has been suggested that canonical structures are more robust because of their more restricted nature compared to the wild and open-ended proof search of instance resolution. However, these restrictions force one into bundled representations, and moreover, their more advanced usage requires significant ingenuity, whereas type class usage is straightforward. Furthermore, wild and open-ended proof search is harmless for predicate classes for which only existence---not identity---matters.

Unification hints are a more general mechanism than type classes, and could provide a more precise account of the interaction between implicit argument inference and proof search. It is not a great stretch to conjecture that a fruitful approach might be to use unification hints as the underlying mechanism, with type classes as an end-user interface encapsulating a particularly convenient idiom for using them.

There are really only two pending concerns that keeps us from making an unequivocal endorsement of type classes as a versatile, expressive, and elegant means of organizing proof developments. The first and lesser of the two is universe polymorphism for definitions as described in the previous section. The second is instance resolution efficiency. In more complex parts of our development we are now experiencing increasingly serious efficiency problems, despite having already made sacrifices by artificially inhibiting many natural class instances in order not to further strain instance resolution.
Fortunately, there is plenty of potential for improvement of the current instance resolution implementation. One source is the vast literature on efficient implementation of Prolog-style resolution, which the hint-based proof search used for instance resolution greatly resembles. We emphasize that these efficiency problems only affect type checking; efficiency of computation using type-checked terms is not affected.

We are currently in the process of retrofitting the rationals interface into CoRN. In future work we aim to base our development of its reals on an abstract dense set, allowing us to use the efficient dyadic rationals~\cite{boldo2009combining} as a base for exact real number computation in Coq~\cite{Oconnor:real,Riemann}. The use of category theory has been important in these developments.

An obvious topic for future research is the extension from equational logic with dependent types~\cite{Cartmell,palmgren2007partial}. Another topic would be to fully, but practically, embrace the
categorical approach to universal algebra~\cite{pitts2001categorical}.

According to \lstinline|coqwc|, our development consists of 5660 lines of specifications and 937 lines of proofs.

\emph{Acknowledgements.}
This research was stimulated by the new possibilities provided by the introduction of type classes
in Coq. In fact, this seems to be the first substantial development that tries to exploit all their
possibilities. As a consequence, we found many small bugs and unintended behavior in the type
class implementation. All these issues were quickly solved by Matthieu Sozeau. Discussions with
Georges Gonthier and Claudio Sacerdoti Coen have helped to
sharpen our understanding of the relation with Canonical Structures and with Unification Hints. Finally, James McKinna suggested many improvements.

This work has been partially funded by the FORMATH project, nr.\ 243847, of the FET program within the 7th Framework program of the European Commission.
\bibliographystyle{plain}
\bibliography{alg}
\end{document}